# Elastic metamaterials for independent realization of negativity in density and stiffness

Joo Hwan Oh[1], Young Eui Kwon[2], Hyung Jin Lee[3] and Yoon Young Kim[1, 3*]

[1]*Institute of Advanced Machinery and Design, Seoul National University,*

*599 Gwanak-ro, Gwanak-gu, Seoul 151-744, Korea*

[2]*Korea Institute of Nuclear Safety, 62 Gwahak-ro, Yuseoung-gu, Daejeon 305-338, Korea*

[3]*School of Mechanical and Aerospace Engineering, Seoul National University,*

*599 Gwanak-ro, Gwanak-gu, Seoul 151-744, Korea*

**Abstract**

In this paper, we present the realization of an elastic metamaterial allowing independent tuning of negative density and stiffness for elastic waves propagating along a designated direction. In electromagnetic (or acoustic) metamaterials, it is now possible to tune permittivity (bulk modulus) and permeability (density) independently. Apparently, the tuning methods seem to be directly applicable for elastic case, but no realization has yet been made due to the unique tensorial physics of elasticity that makes wave motions coupled in a peculiar way. To realize independent tunability, we developed a single-phased elastic metamaterial supported by theoretical analysis and numerical/experimental validations.

---

[*] Corresponding Author, Professor, yykim@snu.ac.kr, phone +82-2-880-7154, fax +82-2-872-1513

Interest in negative material properties has initially grown in electromagnetic field[1-3], being realized by metamaterials. Since then, several interesting applications in antennas, lenses, wave absorbers and others have been discussed. The advances in electromagnetic metamaterials are largely due to independent tuning of the negativity in permittivity and permeability. As done in electromagnetic waves, negative material properties can be also independently tuned in acoustics[4-8] by the analogy between electromagnetic and acoustic waves. Here, our interest is on elastic waves in solid media. We aim to realize solid elastic metamaterials the material properties of which can be independently tuned to be negative for uni-axial wave propagation. In spite of great potentials of single- and double-negative solid metamaterials useful for super vibration shielding[9-11], over-the-diffraction-limit ultrasonic imaging[12-16] and elastic cloaking[17-19], there is no realization to independently tune density and stiffness in solid elastic media.

In that both acoustic and elastic waves propagate by particle motions or oscillations, one may immediately conjecture that the independent tuning method of negativity in material property developed for acoustic waves can be directly used in problems dealing with elastic waves. However, due to the unique tensorial physics of elasticity, complex coupling occurs among various deformation modes such as longitudinal, bending and shear motions. Therefore, the ideas from electromagnetic/acoustic waves cannot be directly used. Earlier investigations[20-27] in elastic metamaterials showed that a monopole or dipole resonance mode of a co-axial internal resonator can be utilized to make bulk modulus or density negative. Among them, Liu *et al.*[25] and Bigoni *et al.*[26] proposed chiral elastic metamaterials consisting of solid materials and Zhu *et al.*[27] realized a single-phased elastic metamaterial. Dubois *et al.*[28,29] realized flat lens and superoscillations with double-negative elastic metamaterials. The limitation of these investigations is, however, the lack of independent tunability of negative stiffness and density because any change in the resonator configuration simultaneously varies the resonance frequencies governing negative stiffness and density. Recently, Lai *et al.*[30] presented the simulation results for metamaterial that independently controls negative stiffness and density by multi-phased resonators. However, the realization of an elastic metamaterial having independent tunability of negative stiffness and density with experimental verification has not been achieved.

Here, we present the realization of an elastic metamaterial allowing independent negativity tuning in stiffness and density. To this end, a single-phased elastic metamaterial in Fig. 1 is developed. As in earlier investigations[20-27], we use resonance modes for the tuning. However, the uniqueness in the present study is that the unit cell of the proposed metamaterial involves two single-phased independent resonators each of which realizes negative density or stiffness. The elastic wave in consideration is the lowest symmetric Lamb wave mode uni-axially propagating along the $x$ direction, i.e., the S0 wave mode the dominant motion of which is in the propagating direction[31]. Referring to Fig. 1, the metamaterial has so-called $x$- and $y$-resonating parts which have local resonance modes in the $x$-direction (propagation direction) and in the $y$-direction (direction perpendicular to the propagation direction), respectively. While more details of the involved physics will be given later, the negative density and stiffness are realized by the $x$- and $y$-resonating parts, respectively. Since the resonance frequencies of the two resonators can be independently tuned, independent tuning of negativity in stiffness and density is available and single or double negativity for a selected range of frequencies can also be achieved. Especially, in realizing negative stiffness by the $y$-resonating parts, the coupling of different deformation modes, longitudinal and bending, is elaborately used; the counterpart of the coupling cannot be found in electromagnetic or acoustic problems. In the present realization of the metamaterial, the inclinations of slender members connected to the central rectangular mass in the $y$-resonating part make the coupling effect which plays a key role for the stiffness tuning.

To illustrate our idea clearly, we will consider the metamaterial in Fig. 1. As Fig. 1 suggests, the square unit cell ($C_{mk}$) of the metamaterial has two resonating parts the physics of which can be better investigated by square unit cells $C_m$ and $C_k$ in Fig. 1. Here, the symbols $C_m$ and $C_k$ denote the unit cell exhibiting negative mass and stiffness, respectively, for the S0 wave mode propagating in the $x$ direction. Because the mechanics of the unit cell $C_{mk}$ can be viewed as a combination of $C_m$ and $C_k$ as illustrated in Fig. 1, how the negativities are realized by $C_m$ and $C_k$ will be investigated first. To model the mechanics of $C_m$ and $C_k$ in association with the wave propagating in the $x$ direction, the elastic metamaterial solids forming $C_m$ and $C_k$ will be replaced by the mass-spring system shown in Figs. 2 (a) and (b). Since we are mainly focused on the S0 wave mode, only particle motion on the $x$-$y$ plane will be considered in the mass-spring system. Although the S0 wave mode is a 3-dimensional wave phenomenon, its characteristics can be analyzed with a two-dimensional model at sufficiently low frequencies[14,15,27].

Firstly, Fig. 2 (a) shows the discrete mass-spring system equivalent to $C_m$. Referring to the mass-spring system in Fig. 2 (a), the unit cell can be considered as the system consists of a lumped mass of $m_1$, an internal mass of $m_3$, a longitudinal spring of $\alpha$ and two bending springs of $\delta$. From the wave analysis in the supplementary material[32], one can find the following effective mass $m_x^{eff}$ and stiffness $\alpha_x^{eff}$ of $C_m$ along the $x$ direction as

$$m_x^{eff} = m_1 + \omega_x^2 m_3 / (\omega_x^2 - \omega^2), \quad \alpha_x^{eff} = \alpha. \tag{1}$$

In equation (1), $\omega$ denotes the excitation angular frequency and $\omega_x = \sqrt{2\delta/m_3}$ corresponds to the resonance frequency of the $x$-resonating part when $m_3$ oscillates in the $x$ direction. If $\omega$ is much lower than $\omega_x$, the effective mass simply becomes the total mass,

-4-

$m_1 + m_3$. However, for $\omega$ near $\omega_x$, the motion of the $x$-resonating part significantly affects the effective mass, making it negative when $\omega$ becomes slightly larger than $\omega_x$. The equation (1) to calculate the effective mass looks similar to the expression obtained earlier with an elastic metamaterial using a coaxial internal resonator[33] if $\omega_x$ is replaced by its dipole resonance frequency. It is worth remarking that in the present case, the motion of the resonating part involves bending deformation of slender members mainly and interacts with the main motion of $m_1$, thus avoiding a need to use multi-phased materials as used to make the coaxial resonators.

Let us now present our method to tune negative stiffness independently from negative mass by using $C_k$. Note that $C_k$ should be so designed that we can tune, independently from $\omega_x$, the resonance frequency to be associated with negative stiffness. Here, our proposition is to utilize coupling between $x$ and $y$ directional motions by intentionally inclining the slender members connecting $m_1$ and $m_2$. The coupled stiffness is denoted by $\gamma$ in Fig. 2 (b) which represents the coupling effect between $x$ and $y$ directional motions. Following the detailed analysis procedure given in the supplementary material[32], the effective mass $m_x^{eff}$ and stiffness $\alpha_x^{eff}$ of $C_k$ along the $x$ direction, which are valid for low frequencies of interest, reduce to

$$m_x^{eff} = m_1 + 2m_2, \quad \alpha_x^{eff} = \alpha - 2\gamma^2 / [m_2(\omega_y^2 - \omega^2)] \tag{2}$$

where $\omega_y = \sqrt{2\beta / m_2}$ is the resonance frequency of the $y$-resonating part as it oscillates in the $y$ direction. Equation (2) shows that the $y$-resonating part affects the effective spring coefficient only while making the effective mass coefficient unaffected by $\omega_y$. Obviously, the effective stiffness $\alpha_x^{eff}$ becomes negative at frequencies slightly lower than $\omega_y$. It is remarked that there were some earlier attempts, without actual realization or experiment, to utilize y-directional motions[21] or a nonlinear phenomenon[34] to control the x-directional stiffness.

Because the unit cell $C_{mk}$ can be viewed as the combination of $C_m$ and $C_k$, the effective mass and stiffness for the metamaterial made of $C_{mk}$ can be written as

$$m_x^{eff} = m_1 + 2m_2 + \omega_x^2 m_3 / (\omega_x^2 - \omega^2), \quad \alpha_x^{eff} = \alpha - 2\gamma^2 / [m_2(\omega_y^2 - \omega^2)]. \tag{3}$$

The discrete version of the unit cell $C_{mk}$ is illustrated in Fig. 2 (c). Equation (3) reveals that our goal to realize the independently-tunable effective mass and stiffness can be indeed achieved by the proposed metamaterials. Because $\omega_x$ and $\omega_y$ are independently tunable, one can realize single- or double-negative metamaterials for different ranges of frequencies. In that the present metamaterial realizes any of $C_{mk}$, $C_m$ and $C_k$ only with a single-phased material, its fabrication is also easy. The validity of equation (3) is also given in the section on 'validation of the effective mass and stiffness' in the supplementary material[32].

Let us now investigate, in some details, the dispersion curves and frequency dependences of x-directional effective density $\rho_x^{eff} = m_x^{eff} / d^2 h$ and modulus of elasticity $C_x^{eff} = \alpha_x^{eff} / h$ where $h$ is the unit cell's thickness and $d$, its width and height. As clearly shown in Figs. 3 (a, b), the metamaterials made of $C_m$ and $C_k$ exhibit single negativities in density and stiffness near $f_x$ ($=\omega_x/2\pi$) and $f_y$ ($=\omega_y/2\pi$), respectively, which are also confirmed by the dispersion curves[35] showing the formation of stop bands.

The sketches of the deformation patterns of $C_m$ and $C_k$ around or near $f_x$ and $f_y$, respectively, show how the negativity in density and stiffness is realized. In Fig. 3 (a), $m_1$, the main mass parts through which the unit cells are connected to each other, moves to the right while $m_3$ moves to the left. Thus, the total momentum of $C_m$ becomes negative because of the 180° out-of-phase motion of $m_3$ for a positive velocity of the unit cell, causing the effective density negative. On the other hand, Fig. 3 (b) shows the mode shape of the unit cell $C_k$ sketched at a frequency just below the resonance frequency $f_y$. Because of the large up- and down-ward y-directional motions of the $m_2$ parts under a force (motion) at the left side of $C_k$, the right side of $C_k$ moves to the left, the opposite direction to the force (motion) at the left side. Thereby, the effective stiffness becomes negative. This is possible because of the elaborate coupling of *x* and *y* directional deformations which cannot be found in electromagnetic or acoustic wave cases.

The metamaterials composed of $C_{mk}$, which is the combination of $C_m$ and $C_k$, exhibit the combined effects of the two independent metamaterials made of $C_m$ and $C_k$, as clearly demonstrated in Fig. 3 (c). This finding indeed confirms the independent tunability of the proposed metamaterial in its density and stiffness values. As being obvious, there appears a passing zone with the negative group velocity (i.e. the negative slope in the dispersion curve) in the zone of overlapping frequencies of the negative density in $C_m$ and the negative stiffness in $C_k$. Because $f_x$ and $f_y$ can be independently tuned in the developed metamaterial, the metamaterials can be tailored to meet specific applications requiring single and/or double negativity.

Finally, the metamaterials are fabricated and their wave characteristics are experimentally investigated. Fig. 4 illustrates the experimental setup and also shows a sample of transmitted and measured signals for the experiments. Three sets of experiments were performed with the metamaterials consisting of $C_k$, $C_m$ and $C_{mk}$, which are made of an aluminum only. Since these metamaterials are made of a single-phased low-loss aluminum, adverse effects of loss can be insignificant. The S0 wave is actuated by the actuating piezoelectric transducers with the modulated Gaussian pulses centered at 15, 25 and 35 kHz. Since the metamaterials are dispersive[36], the measured signals were analyzed through the Short-Time Fourier Transform (STFT). Fig. 5 shows the experimental results; the detailed experimental procedures, the process of STFT and meaning of the color level in Fig. 5 can be found in the supplementary material[32].

Fig. 5 shows that the elastic metamaterial having only the *x*- or *y*-resonating part exhibits no wave transmission in the frequency ranges in which either one of the effective density and stiffness is negative; see Figs. 5 (a, b). At the excitation frequency of 25 kHz lying between $f_x$ and $f_y$, the wave is transmitted through the metamaterial made of $C_{mk}$ as confirmed in Fig. 5 (c). To further investigate the transmitted waves, analytically calculated arrival times $t_a$ are plotted by the white dotted line in Fig. 5 (c). The comparison of the experimentally- and analytically-calculated arrival times confirms the validity of the wave dispersion relations of the proposed metamaterial. At around 27 kHz, however, little transmission occurs because of a large impedance mismatch between the metamaterial and the aluminum plate. Also, the second arrival pulse component of 25 kHz appears around 1.5 ms due to internal reflections within the metamaterials and its appearance cannot be predicted by the dispersion analysis. These results show that the independent achievement of negative stiffness and density is realized by the proposed metamaterials. More experimental results showing the independent tunability can be found in the supplementary material[32].

This study presents the metamaterial realization with independent tuning of effective negative density and stiffness with a single-phased material. In spite of apparent similarity between elastic waves and electromagnetic/acoustic waves, the independent negativity tuning in elastic metamaterials has not been realized earlier. To realize the independent negativity tuning, a single-phased elastic metamaterial was proposed. Here, among others, the independent-tunable negative stiffness was realized by a locally-resonating part the motion of which is dominant in the direction perpendicular to the wave propagation direction. The theoretical and experimental wave analyses of the metamaterials were carried out to confirm the independent tunability of negativity, with realizations of single negative density/stiffness and simultaneous negativity in density and stiffness. Considering many practical important applications of elastic waves in ultrasonic imaging, vibration shielding, etc., our metamaterials could lead to active explorations in elastic metamaterials, which currently appear to be less active in the community.

**Acknowledgements**

This research was supported by the National Research Foundation of Korea (NRF) funded by the Ministry of Science, ICT & Future Planning, Korea (Nos: 2015021967 and CAMM-2014M3A6B3063711) contracted through IAMD at Seoul National University.

**Figure Captions**

Fig. 1. (Color online) The sketch of the proposed unit cell $C_{mk}$ which can be constructed by two independent unit cells $C_m$ and $C_k$ realizing negative effective mass and stiffness along the *x* direction, respectively.

Fig. 2. (Color online) The mass-spring unit cell model corresponding to (a) the unit cell $C_m$ for negative density, (b) the unit cell $C_k$ for negative stiffness and (c) the unit cell $C_{mk}$ for independent tuning of negativity in mass and stiffness.

Fig. 3. (Color online) The *x*-directional dispersion curve, effective parameters and mode shapes of the proposed unit cells (a) $C_m$ (b) $C_k$ and (c) $C_{mk}$.

Fig. 4. (Color online) Experimental setting for wave propagation through the developed metamaterials with a photo of the fabricated metamaterial. The detailed geometries of the unit cells are given in Fig. S3.

Fig. 5. (Color online) Comparison between experimental and analytic results ($k_x$: wave number in the *x* direction, $t_a$: arrival time) for the metamaterials consisting of (a) $C_m$ (b) $C_k$ and (c) $C_{mk}$. Here, DP, SN and DN stand for double positivity, single negativity and double negativity, respectively. In (c), the white dotted-line denotes the analytically calculated arrival time.

**Figures**

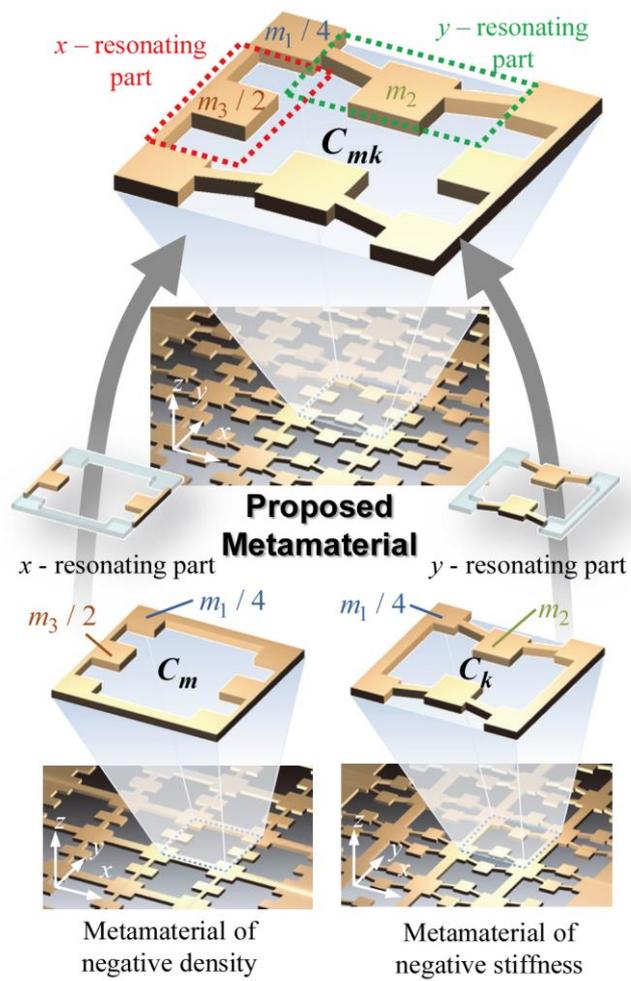

Fig. 1.

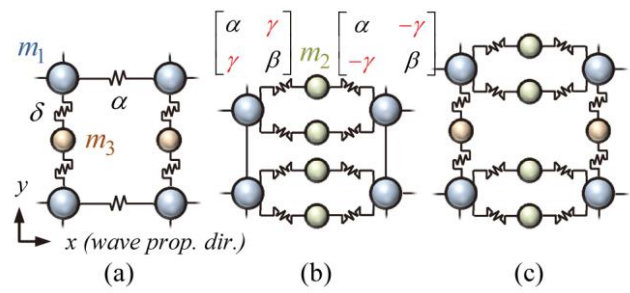

Fig. 2.

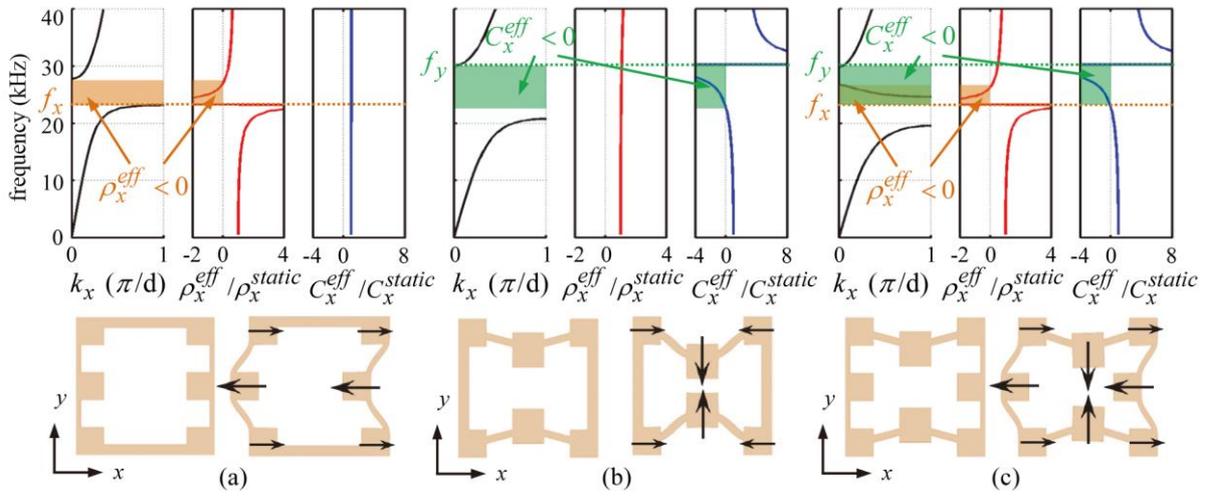

Fig. 3.

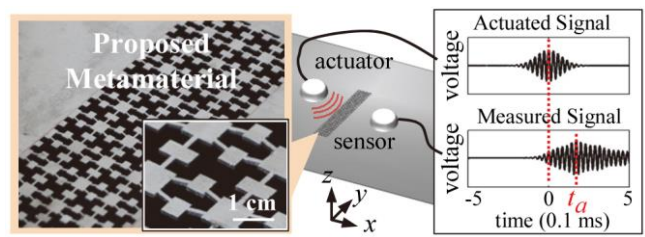

Fig. 4.

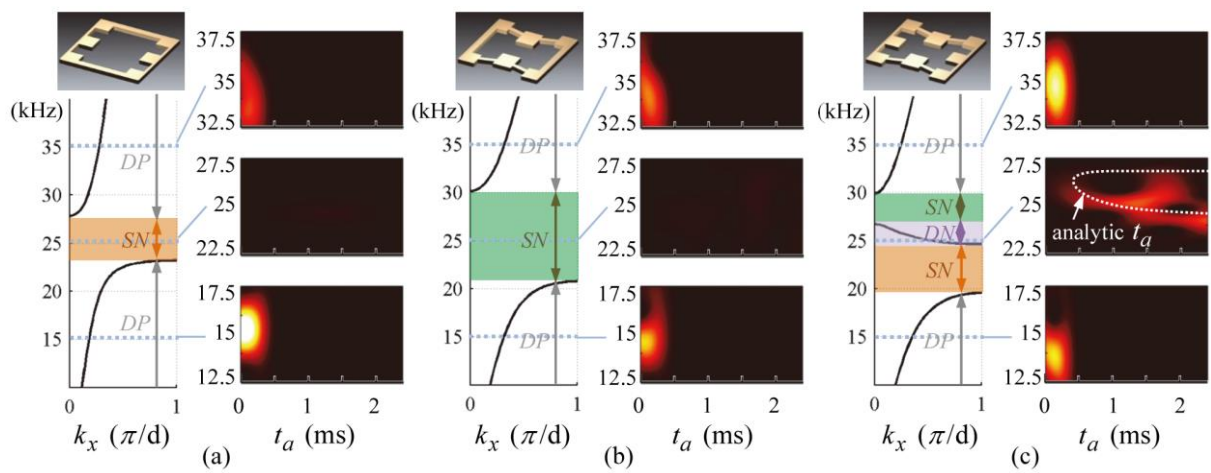

Fig. 5.

# Supplementary Materials

# Elastic metamaterials for independent realization of negativity in density and stiffness

Joo Hwan Oh[1], Young Eui Kwon[2], Hyung Jin Lee[3] and Yoon Young Kim[1, 3†]

[1]*Institute of Advanced Machinery and Design, Seoul National University,*

*599 Gwanak-ro, Gwanak-gu, Seoul 151-744, Korea*

[2]*Korea Institute of Nuclear Safety, 62 Gwahak-ro, Yuseoung-gu, Daejeon 305-338, Korea*

[3]*School of Mechanical and Aerospace Engineering, Seoul National University,*

*599 Gwanak-ro, Gwanak-gu, Seoul 151-744, Korea*

The *x*-directional effective mass and stiffness of the metamaterials formed by the unit cells $C_m$, $C_k$ and $C_{mk}$ will be analytically derived by using equivalent discrete mass-spring systems. The equivalence between the discrete model and the original continuum model will be verified through the dispersion curves. The effective mass and stiffness are numerically estimated by using a retrieval method of effective material properties applied to the original solid model. Also, the experimental setup and procedure are explained in some details.

**Background: basic equations for a simple periodic mass-spring system**

To begin with, we review the well-known dispersion equation and expression of characteristic impedance in a simple one-dimensional periodic mass-spring system in Fig. S1. The system consists of an infinite number of periodically-arranged lumped masses of mass $m_x^{eff}$ and springs of stiffness $\alpha_x^{eff}$. Starting from the forces acting on $m_x^{eff}$, the equation of motion can be derived as[S1]

$$m_x^{eff} \frac{\partial^2 u_n}{\partial t^2} = \alpha_x^{eff}(u_{n+1} - u_n) - \alpha_x^{eff}(u_n - u_{n-1}). \quad (S1)$$

where $u_n$ is the *x*-directional displacement of the $n^{\text{th}}$ mass and *t* denotes time. Assuming time harmonic wave motion at an angular frequency of $\omega$, $u_n$ can be expressed as $u_n = U_n \exp[i(\omega t - kx)]$ ($i = \sqrt{-1}$) where *k* denotes the wavenumber. Noting that $u_{n+1} = \exp(-ikd)u_n$ and $u_{n-1} = \exp(ikd)u_n$, equation (S1) yields the following dispersion equation,

$$-\omega^2 m_x^{eff} = \alpha_x^{eff}(\exp(-ikd) + \exp(ikd) - 2). \quad (S2)$$

---

[†] Corresponding Author, Professor, yykim@snu.ac.kr, phone +82-2-880-7154, fax +82-2-872-1513

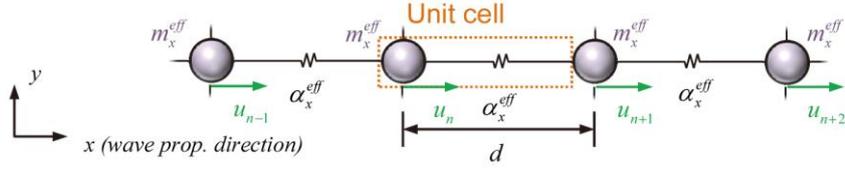

**Fig. S1.** A periodic mass-spring system under uni-axial wave motion in the *x*-direction.

To uniquely define the effective mass and stiffness from the dispersion relation, the expression for characteristic impedance is also needed. Using the force *F* exerted on the $(n+1)^{\text{th}}$ mass by the $n^{\text{th}}$ mass,

$$F = \alpha_x^{eff}(u_n - u_{n+1}) = \alpha_x^{eff}(\exp(ikd) - 1)u_{n+1}, \quad (S3)$$

the characteristic impedance *Z* is written as[S1]

$$Z = \frac{\sigma_{xx}}{\partial u_{n+1}/\partial t} = \frac{F/d}{\partial u_{n+1}/\partial t} = \frac{F}{i\omega d u_{n+1}} = -\frac{Fi}{\omega d u_{n+1}} = \alpha_x^{eff} \frac{(1-\exp(ikd))i}{\omega d}. \quad (S4)$$

As can be seen in equation (S4), one can determine the effective stiffness of a metamaterial system by writing its characteristic impedance equation and comparing the result with equation (S4). Then, the effective mass can be determined by comparing the dispersion equation of a system of interest with equation (S2).

**Derivation of the effective mass and stiffness of the metamaterials in Fig. 1**

As remarked earlier, we should derive the dispersion curves and impedance characteristics of systems of interest in forms similar to equations (S2, 4). To this end, we will use the discrete models of the three unit cells, $C_m$, $C_k$ and $C_{mk}$ shown in Fig. 2. The discrete models are re-plotted in Fig. S2 with illustrations of displacements of masses $m_1$, $m_2$ and $m_3$. As explained in the manuscript, although the S0 wave mode mainly considered in this work is a 3-dimensional wave phenomenon, it can be accurately analyzed with a two-dimensional mass-spring system if the operating frequency range is sufficiently low as in this work. Starting from Fig. S2 (a) representing $C_m$, the x-directional displacements of $m_1$ and $m_3$ in the $(n, j)^{th}$ unit cell are denoted by $u_{n,j}^1$ and $u_{n,j}^3$, respectively. The indices $n$ and $j$ denote the unit cell location in the $x$ and $y$ coordinates. Considering the $x$-directional displacement of $m_1$ to analyze the S0 wave mode propagating along the $x$ direction, we use the following equations of motion:

$$m_1 \frac{\partial^2 u^1_{n,j}}{\partial t^2} = \alpha(u^1_{n+1,j} + u^1_{n-1,j} - 2u^1_{n,j}) + \delta(u^3_{n,j+1} + u^3_{n,j} - 2u^1_{n,j}), \quad \text{(S5a)}$$

$$m_3 \frac{\partial^2 u^3_{n,j}}{\partial t^2} = \delta(u^1_{n,j} + u^1_{n,j-1} - 2u^3_{n,j}). \quad \text{(S5b)}$$

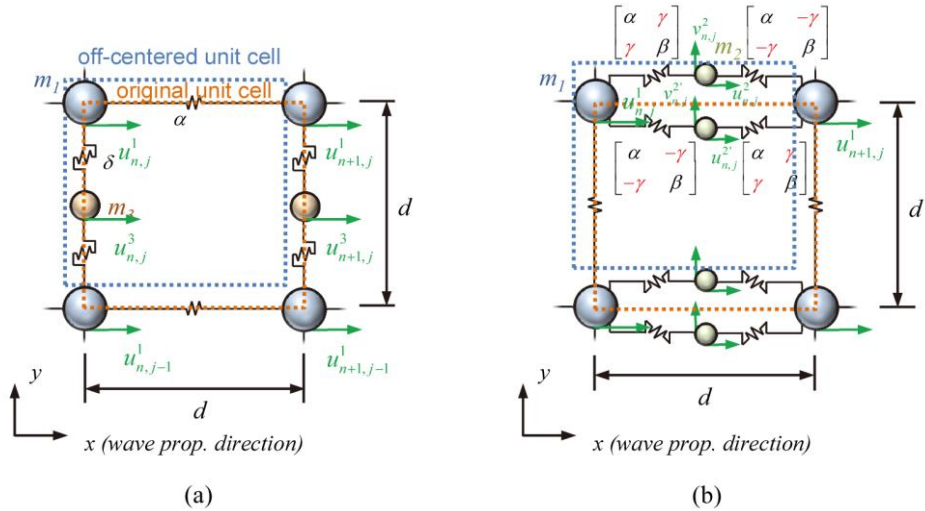

**Fig. S2.** The discrete mass-spring systems corresponding to the elastic metamaterials made of (a) $C_m$ and (b) $C_k$.

Note that there is no need to consider displacements in the $y$ axis because the $y$-displacements do not interact with the S0 wave mode (dominated by $x$-displacements of $m_1$). Also, the displacements $u^1_{n,j}$ and $u^3_{n,j}$ can be assumed not to vary along the $y$ axis. Therefore, one can set $u^3_{n,j+1} = u^3_{n,j}$ and $u^1_{n,j+1} = u^1_{n,j}$.

Assuming time-harmonic wave motion at an angular frequency of $\omega$, equation (S5) can be re-written as

$$-\omega^2 m_1 u^1_{n,j} = \alpha(\exp(ikd) + \exp(-ikd) - 2)u^1_{n,j} + 2\delta(u^3_{n,j} - u^1_{n,j}), \quad \text{(S6a)}$$

$$-\omega^2 m_3 u^3_{n,j} = 2\delta(u^1_{n,j} - u^3_{n,j}). \quad \text{(S6b)}$$

From equation (S6b),

$$u^3_{n,j} = \frac{2\delta}{2\delta - \omega^2 m_3} u^1_{n,j}. \quad \text{(S7)}$$

Substituting equation (S7) into equation (S6a) yields

$$-\omega^2 m_1 u^1_{n,j} = \alpha(\exp(ikd) + \exp(-ikd) - 2)u^1_{n,j} + \frac{2\delta m_3 \omega^2}{2\delta - \omega^2 m_3} u^1_{n,j}. \quad \text{(S8)}$$

From (S8), the following dispersion relation can be obtained for the system made of $C_m$:

$$-\omega^2 (m_1 + \frac{2\delta m_3}{2\delta - \omega^2 m_3}) = \alpha(\exp(ikd) + \exp(-ikd) - 2). \quad \text{(S9)}$$

To find an expression for the characteristic impedance, we write the force exerted on the $(n+1)^{th}$ mass by the $n^{th}$ mass as

$$F = \alpha(u_{n,j}^1 - u_{n+1,j}^1) = \alpha(\exp(ikd)-1)u_{n+1,j}^1. \tag{S10}$$

Using equation (S10), one can identify the characteristic impedance as

$$Z = -\frac{Fi}{\omega d u_{n+1,j}^1} = \alpha \frac{(1-\exp(ikd))i}{\omega d}. \tag{S11}$$

Comparing equations (S9, 11) with equations (S2, 4), the effective mass and stiffness can be identified as

$$m_x^{eff} = m_1 + \frac{2\delta m_3}{2\delta - \omega^2 m_3} = m_1 + \frac{\omega_x^2 m_3}{\omega_x^2 - \omega^2}, \tag{S12a}$$

$$\alpha_x^{eff} = \alpha, \tag{S12b}$$

where $\omega_x = \sqrt{2\delta/m_3}$ is the resonance frequency of the x-resonating part.

Let us now derive the effective mass and stiffness for the metamaterial made of $C_k$ by using the discrete model in Fig. S2 (b). Here, we should pay a special attention to the coupled stiffness connecting $m_1$ and $m_2$, which can be expressed in the following relations:

$$\begin{Bmatrix} F_x \\ F_y \end{Bmatrix} = \begin{bmatrix} \alpha & \gamma \\ \gamma & \beta \end{bmatrix} \begin{Bmatrix} u \\ v \end{Bmatrix} \text{ or } \begin{Bmatrix} F_x \\ F_y \end{Bmatrix} = \begin{bmatrix} \alpha & -\gamma \\ -\gamma & \beta \end{bmatrix} \begin{Bmatrix} u \\ v \end{Bmatrix}, \tag{S13}$$

where $u$ and $v$ denotes the x- and y-directional displacements applied to the coupled spring. The off-diagonal term $\gamma$ appears in equation (S13) due to the coupling between x- and y-displacements that is intentionally realized by the inclinations of the slender members connecting $m_1$ and $m_2$. Referring to the original continuum configuration of $C_k$ shown in Fig. 1, one can see why the $-\gamma$ term also appears; the two symmetric members with the opposite inclination angles are used to connect $m_1$ and $m_2$. As a result, not only the x-directional but also the y-directional motions of $m_2$ should be considered in deriving equations of motion.

Writing up all equations of motion for $m_1$ and $m_2$,

$$m_1 \frac{\partial^2 u_{n,j}^1}{\partial t^2} = -4\alpha u_{n,j}^1 + \alpha u_{n,j}^2 + \alpha u_{n-1,j}^2 + \alpha u_{n,j}^{2'} + \alpha u_{n-1,j}^{2'} + \gamma v_{n,j}^2 - \gamma v_{n-1,j}^2 - \gamma v_{n,j}^{2'} + \gamma v_{n-1,j}^{2'}, \quad \text{(S14a)}$$

$$m_2 \frac{\partial^2 u_{n,j}^2}{\partial t^2} = -2\alpha u_{n,j}^2 + \alpha u_{n+1,j}^1 + \alpha u_{n,j}^1, \quad \text{(S14b)}$$

$$m_2 \frac{\partial^2 u_{n,j}^{2'}}{\partial t^2} = -2\alpha u_{n,j}^{2'} + \alpha u_{n+1,j}^1 + \alpha u_{n,j}^1, \quad \text{(S14c)}$$

$$m_2 \frac{\partial^2 v_{n,j}^2}{\partial t^2} = -\gamma u_{n+1,j}^1 + \gamma u_{n,j}^1 - 2\beta v_{n,j}^2, \quad \text{(S14d)}$$

$$m_2 \frac{\partial^2 v_{n,j}^{2'}}{\partial t^2} = \gamma u_{n+1,j}^1 - \gamma u_{n,j}^1 - 2\beta v_{n,j}^{2'}. \quad \text{(S14e)}$$

In equation (S14), the quantities with superscripts 2 and $2'$ are related to the upper and the lower mass $m_2$, respectively. Assuming time-harmonic wave motion, equations (S14) can be re-written as

$$\begin{aligned}-\omega^2 m_1 u_{n,j}^1 &= -4\alpha u_{n,j}^1 + \alpha(1+\exp(ikd))u_{n,j}^2 + \alpha(1+\exp(ikd))u_{n,j}^{2'} \\ &\quad + \gamma(1-\exp(ikd))v_{n,j}^2 - \gamma(1-\exp(ikd))v_{n,j}^{2'}\end{aligned}, \quad \text{(S15a)}$$

$$-\omega^2 m_2 u_{n,j}^2 = \alpha(1+\exp(-ikd))u_{n,j}^1 - 2\alpha u_{n,j}^2, \quad \text{(S15b)}$$

$$-\omega^2 m_2 u_{n,j}^{2'} = \alpha(1+\exp(-ikd))u_{n,j}^1 - 2\alpha u_{n,j}^{2'}, \quad \text{(S15c)}$$

$$-\omega^2 m_2 v_{n,j}^2 = \gamma(1-\exp(-ikd))u_{n,j}^1 - 2\beta v_{n,j}^2, \quad \text{(S15d)}$$

$$-\omega^2 m_2 v_{n,j}^{2'} = -\gamma(1-\exp(-ikd))u_{n,j}^1 - 2\beta v_{n,j}^{2'}. \quad \text{(S15e)}$$

Re-writing equations (S15 b-e) with respect to the displacement $u_{n,j}^1$ yields

$$u_{n,j}^2 = u_{n,j}^{2'} = \frac{\alpha(1+\exp(-ikd))}{2\alpha - \omega^2 m_2} u_{n,j}^1 = \frac{\alpha(\exp(ikd)+1)}{2\alpha - \omega^2 m_2} u_{n+1,j}^1, \quad \text{(S16a)}$$

$$v_{n,j}^2 = -v_{n,j}^{2'} = \frac{\gamma(1-\exp(-ikd))}{2\beta - \omega^2 m_2} u_{n,j}^1 = \frac{\gamma(\exp(ikd)-1)}{2\beta - \omega^2 m_2} u_{n+1,j}^1, \tag{S16b}$$

By substituting equations (S16a, b) to equation (S15a), one can finally obtain the dispersion equation as

$$-\omega^2 (m_1 + \frac{4\alpha m_2}{2\alpha - \omega^2 m_2}) = \left[\frac{2\alpha^2}{2\alpha - \omega^2 m_2} - \frac{2\gamma^2}{2\beta - \omega^2 m_2}\right](\exp(-ikd) + \exp(ikd) - 2). \tag{S17}$$

If the operating frequency of interest is assumed to be much lower than $2\alpha/m_2$ (this situation is very typical because $\alpha$ is usually one order larger than $\beta$ as in Table S1), one can assume that

$$2\alpha - \omega^2 m_2 \approx 2\alpha \tag{S18}$$

and equation (S18) can be simplified to

$$-\omega^2 (m_1 + 2m_2) = \left[\alpha - \frac{2\gamma^2}{2\beta - \omega^2 m_2}\right](\exp(-ikd) + \exp(ikd) - 2). \tag{S19}$$

To find the characteristic impedance for the system, the force exerted on the $(n+1)^{th}$ mass by the $n^{th}$ mass is obtained as

$$F = -(2\alpha u_{n+1,j}^1 - \alpha u_{n,j}^2 - \alpha u_{n,j}^{2'} + \gamma v_{n,j}^2 - \gamma v_{n,j}^{2'}). \tag{S20}$$

Substituting equations (S16a, b) into equation (S20) yields

$$F = \left[-2\alpha + \frac{2\alpha^2(1+\exp(ikd))}{2\alpha - \omega^2 m_2} - \frac{2\gamma^2(\exp(ikd)-1)}{2\beta - \omega^2 m_2}\right] u_{n+1,j}^1. \tag{S21}$$

Using the assumption in equation (S18), equation (S21) is simplified to

$$\begin{aligned} F &= \left[-2\alpha + \alpha(1+\exp(ikd)) - \frac{2\gamma^2(\exp(ikd)-1)}{2\beta - \omega^2 m_2}\right] u_{n+1,j}^1 \\ &= \left[\alpha - \frac{2\gamma^2}{2\beta - \omega^2 m_2}\right](\exp(ikd) - 1) u_{n+1,j}^1 \end{aligned} \tag{S22}$$

Accordingly, the characteristic impedance becomes

$$Z = -\frac{Fi}{\omega d u_{n+1,j}^1} = \left[\alpha - \frac{2\gamma^2}{2\beta - \omega^2 m_2}\right]\frac{(1-\exp(ikd))i}{\omega d}. \tag{S23}$$

Finally, the effective mass and stiffness are identified as, by comparing equations (S19, 23) with equations (S2, 4),

$$m_x^{eff} = m_1 + 2m_2, \tag{S24a}$$

$$\alpha_x^{eff} = \alpha - \frac{2\gamma^2}{2\beta - \omega^2 m_2} = \alpha - \frac{2\gamma^2/m_2}{\omega_y^2 - \omega^2}. \tag{S24b}$$

where $\omega_y = \sqrt{2\beta/m_2}$ is the resonance frequency of the y-resonating part.

Now, the effective mass and stiffness for the elastic metamaterials made of $C_m$ and $C_k$ are identified. Because the metamaterials made of $C_{mk}$ inherit the characteristics of $C_m$ and $C_k$ based metamaterials, the effective mass and stiffness of the metamaterials become

$$m_x^{eff} = m_1 + 2m_2 + \frac{\omega_x^2 m_3}{\omega_x^2 - \omega^2}, \tag{S25a}$$

$$\alpha_x^{eff} = \alpha - \frac{2\gamma^2/m_2}{\omega_y^2 - \omega^2}. \tag{S25b}$$

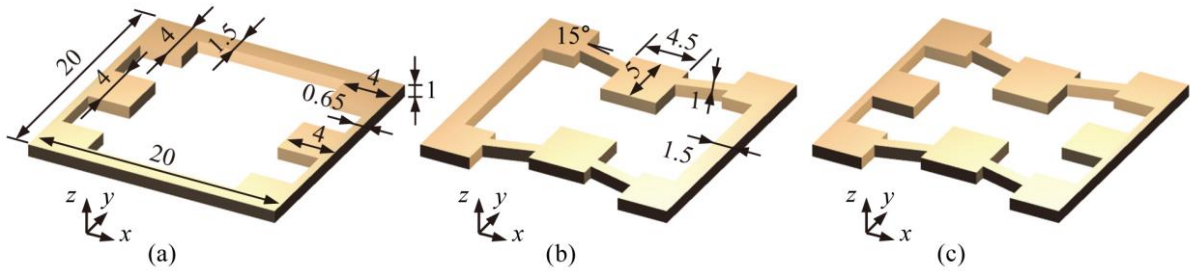

**Fig. S3.** Sketches of the elastic solid unit cells of (a) $C_m$ (b) $C_k$ and (c) $C_{mk}$ with the specific dimensions. The numbers in the figure are all in the millimeter scale.

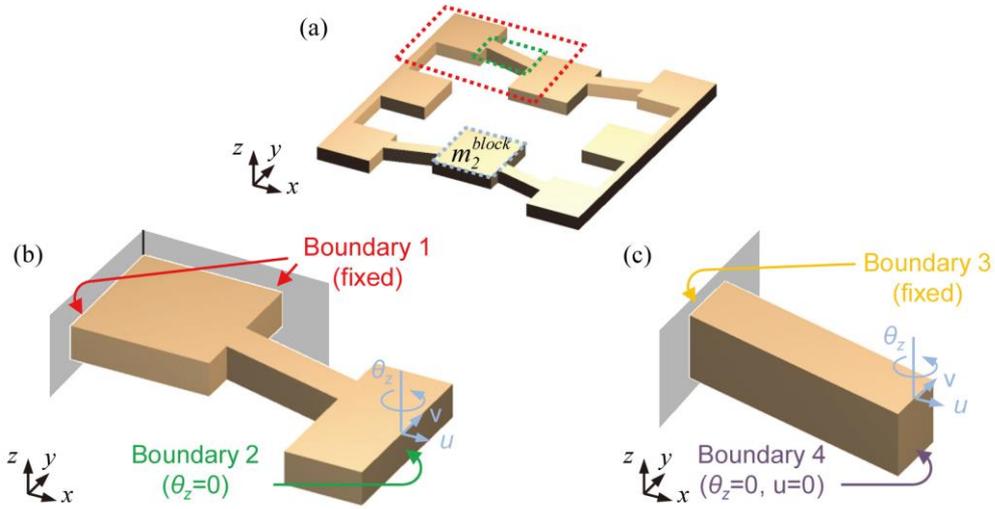

**Fig. S4.** (a) The unit cell of $C_{mk}$, the finite element model to calculate (b) the spring coefficients $\alpha$, $\beta$ and $\gamma$, (c) the additional mass for the mass coefficient $m_2$, for the unit cell of $C_{mk}$.

**Validation of the effective mass and stiffness**

The validation of the effective mass and stiffness previously derived can be made directly or indirectly. As an indirect way, one can compute the dispersion curve by using the one-dimensional dispersion equation (S2) with the effective mass and stiffness in equations (S12,24,25) and compare the curve with that obtained for the original solid unit cell shown in Fig. 1. For actual calculations, the specific dimensions and geometries are given in Fig. S3. To extract the lumped parameters, the mass and spring coefficient of the continuum structures in Fig. S3, one may attempt to derive formula to relate the geometric parameters to the mass and spring coefficients as done in Refs. [S2-S4]. Because of the complexity in geometry, we used the finite element method for accurate estimation. Table 1 lists the calculated mass and stiffness values for the three types of unit cells, which were calculated with 3-dimensional finite element simulations. For instance, the equivalent spring coefficients $\alpha$, $\beta$ and $\gamma$, and the mass coefficient $m_2$, for the unit cell of $C_{mk}$, are calculated with the finite element

|   | Metamaterial with x-resonating part | Metamaterial with y-resonating part | Metamaterials with x- and y-resonating parts |
|---|---|---|---|
| $\alpha$ | 1.24e4 kN/m | 1.02e4 kN/m | 1.02e4 kN/m |
| $\beta$ | 1.15e3 kN/m | 1.23e3 kN/m | 1.23e3 kN/m |
| $\gamma$ | 1.10e-3 kN/m | 2.44e3 kN/m | 2.43e3 kN/m |
| $\delta$ | 1.04e3 kN/m | 3.26e3 kN/m | 1.04e3 kN/m |
| $m_1$ | 2.52e-4 kg | 2.53e-4 kg | 2.21e-4 kg |
| $m_2$ | 1.85e-5 kg | 6.71e-5 kg | 6.71e-5 kg |
| $m_3$ | 9.59e-5 kg | 4.70e-5 kg | 9.59e-5 kg |

**Table S1.** The estimated values of mass and stiffness elements of the discrete mass-spring models corresponding to the unit cells shown in Fig. S3. The calculations were performed by the finite element analysis for accurate evaluation.

model shown in Fig. S4. We will briefly explain how these coefficients are calculated.

To calculate the spring coefficients $\alpha$, $\beta$ and $\gamma$, only the part marked as the red box in Fig. S4 (a) needs to be considered. Here, the fixed boundary conditions are imposed on Boundary 1 and $\theta_z = 0$ (no rotation about the z axis) is imposed on Boundary 2. These conditions are illustrated in Fig. S4 (b). The part in Fig. S4 (b) is discretized by the finite elements and we calculated $u$ and $v$, the x- and y- directional displacements at the center point of Boundary 2 by applying $F_x$ and $F_y$, the x- and y- directional forces, independently. The calculated displacements are used to construct the compliance matrix $\mathbf{B}$ that relates the forces and displacements as $\mathbf{B}\begin{bmatrix} F_x \\ F_x \end{bmatrix} = \begin{bmatrix} u \\ v \end{bmatrix}$. Finally, the spring coefficients $\alpha$, $\beta$ and $\gamma$ can be evaluated from the components of the stiffness matrix by taking the inverse of $\mathbf{B}$, resulting in

$$\begin{bmatrix} \alpha & -\gamma \\ -\gamma & \beta \end{bmatrix} = \mathbf{B}^{-1}. \tag{S26}$$

To calculate the mass coefficient $m_2$, one can simply choose $m_2 = m_2^{block}$ where $m_2^{block}$

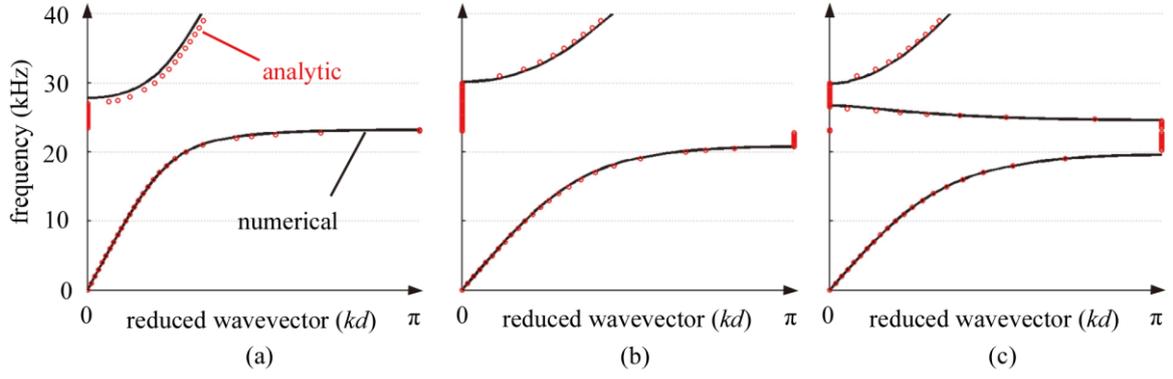

**Fig. S5.** Comparison of the dispersion curves numerically calculated from the original elastic solid metamaterials (in black solid lines) and from the effective mass and stiffness values (in red circles) calculated in by equation (3), equivalently, equations (S12, 24, 25). The results are for the metamaterials made of (a) $C_m$ (b) $C_k$ and (c) $C_{mk}$.

denotes the mass of the block marked as the blue box in Fig. S4 (a). For more precise estimation, the mass of the slender beam part, marked as the green box in Fig. S4 (a), can be additionally considered because the beam part serves not only as stiffness but also as mass[S4]. The beam part in consideration is illustrated as Fig. S4 (c). To accurately evaluate the additional mass coefficient from the slender beam structure, which will be denoted as $m_2^{beam}$, the most accurate method is to perform the free vibration analysis, yielding the lowest bending eigenfrequency of $\omega_{bending}^{beam}$. Because the bending stiffness of the beam segment $\beta^{beam}$ can be evaluated by applying a force in the $y$ direction at the center point of Boundary 4, one can evaluate $m_2^{beam}$ from the well-known eigenfrequency of a single degree-of-freedom vibration as $m_2^{beam} = \beta^{beam} / (\omega_{bending}^{beam})^2$. Finally, we find $m_2^{beam} = \beta^{beam} / (\omega_{bending}^{beam})^2$. If $m_2^{beam}$ is smaller than one tenth of $m_2^{block}$, $m_2^{beam}$ can be actually ignored.

Fig. S5 compares the dispersion curves where the dispersion curves for the original solid unit cells were computed by the 3-dimensional finite element method[S5]. Excellent agreements between the two results are observed for all three cases corresponding to $C_m$, $C_k$ and $C_{mk}$, validating the derived effective mass and stiffness. This clearly shows that the S0 wave mode which is a 3-dimensional wave phenomenon can be accurately described with the proposed two-dimensional mass-spring system.

A more direct way to validate the derived effective mass and stiffness is to estimate the effective parameters by the retrieval method developed for elastic metamaterials[S6]. The details of the retrieval method will not be given here, but the reflection and transmission coefficients of a metamaterial layer consisting of one of the metamaterials in Fig. S3 were numerically calculated with 3-dimensional finite element simulations. Fig. S6 compares the retrieved values and those obtained by equations (S12, 24, 25). For all cases, good agreements were found. Also, from Figs. S6 (a) and (b), the derived expressions for the effective mass and stiffness can be validated.

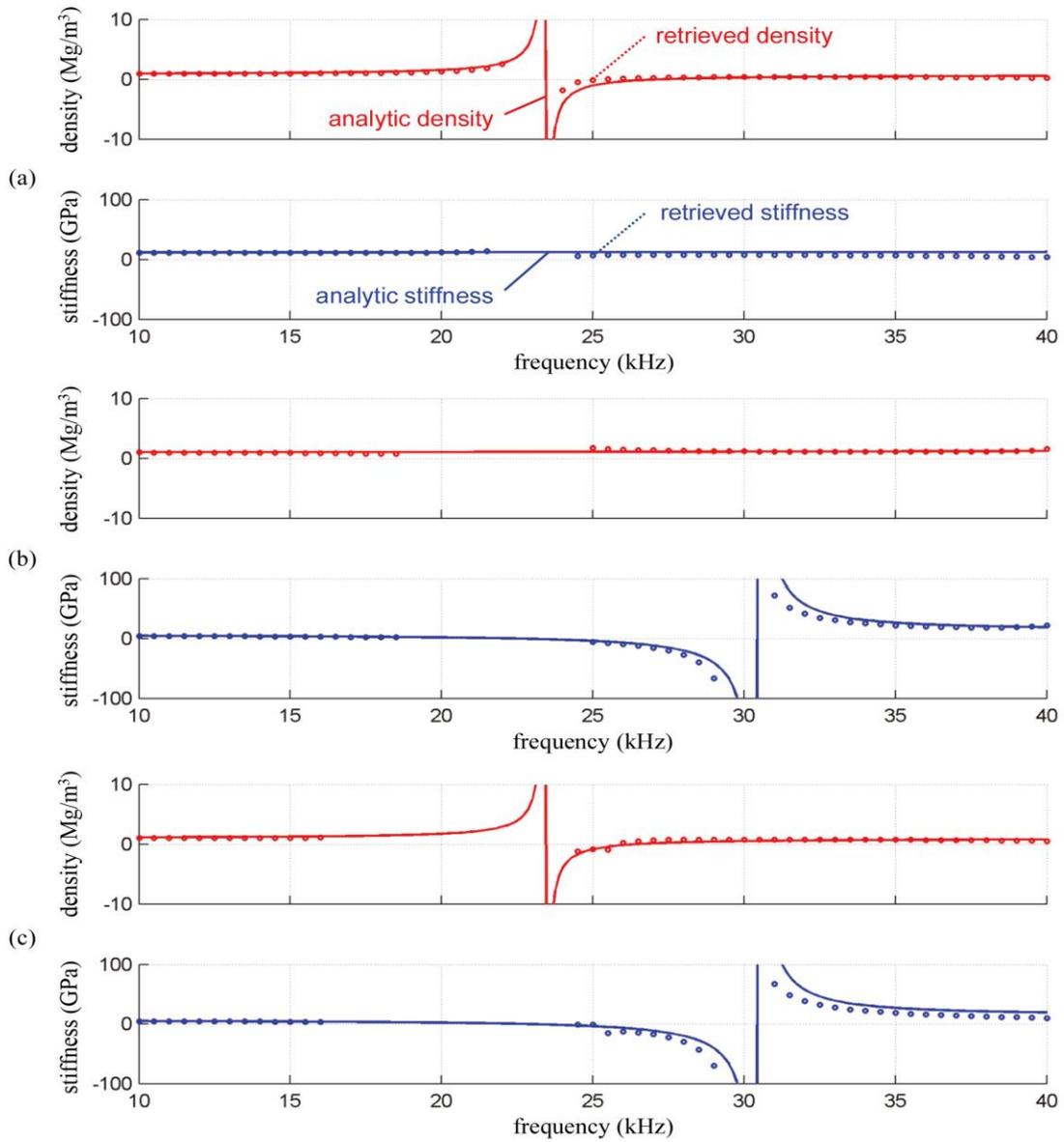

**Fig. S6.** Comparison of the effective mass and stiffness values numerically calculated by the retrieval method applied to the original continuum metamaterials (in circles) and those by equation (3), equivalently, equations (S12, 24, 25) (in solid lines). The results are for the metamaterials made of (a) $C_m$ (b) $C_k$ and (c) $C_{mk}$.

**Numerical simulation to investigate phase velocities in the metamaterials**

Fig. S7 (a) shows the model used to investigate phase velocities inside the metamaterial layer. The metamaterial layer consisting of 4 $C_{mk}$ unit cells is inserted between two aluminum plates. The plates and layer are discretized by finite elements for numerical analysis. For the simulations, the S0 sinusoidal waves of 15, 25 and 35 kHz are incident from the left aluminum plate and the $x$-directional displacements $u_j$ ( $j$ =A, B and C) are measured where A, B and C denote points inside the metamaterial layer. The theory predicts that the metamaterials made of $C_{mk}$ should exhibit positive phase velocities at 15 and 35 kHz, but a negative phase velocity at 25 kHz.

Figs. S7 (b-d) show the simulation results for the displacements $u_A$, $u_B$ and $u_C$ at 15, 25 and 35 kHz, respectively. At 15 and 35 kHz, the crests move forward along the positive $x$ direction as the wave propagates. At 25 kHz, on the other hand, the crests move backwards along the negative $x$ direction as the wave propagates along the positive $x$ direction. This means that the phase velocity at 25 kHz is negative. The phase velocities measured from the numerical simulations are found to be 1739, -2500 and 3333 m/s while the phase velocities calculated from the dispersion curve are 1729, -2353 and 3294 m/s for the frequency of 15, 25 and 35 kHz, respectively. The numerical simulations confirm the validity of the analytic prediction, including the formation of the negative phase velocity at 25 kHz.

It will be interesting to show experimentally the formation of the negativity phase velocity at 25 kHz, the unique phenomenon due to the simultaneous negative effective density and stiffness. Fig. S8 shows the measure displacement fields of $u_A$, $u_B$ and $u_C$. For the experiment, a thin highly-reflective rectangular film was vertically installed at the measurement locations. Then a laser vibrometer (OFV-551, Polytec) was used to measure the $x$-directional displacement. The detailed actuation process will be explained in the next section. Fig. S8 clearly shows the negative phase velocity field and the phase pattern matches well with that obtained in Fig. S7 (c) from the numerical simulation. The magnitude difference between the displacements from the simulation and the experiment is due to the difficulty to install the thin film exactly vertically. Nevertheless, the experimental measurements clearly reveal that the phase velocity in the metamaterial at 25 kHz that belongs to the double negative zone is negative.

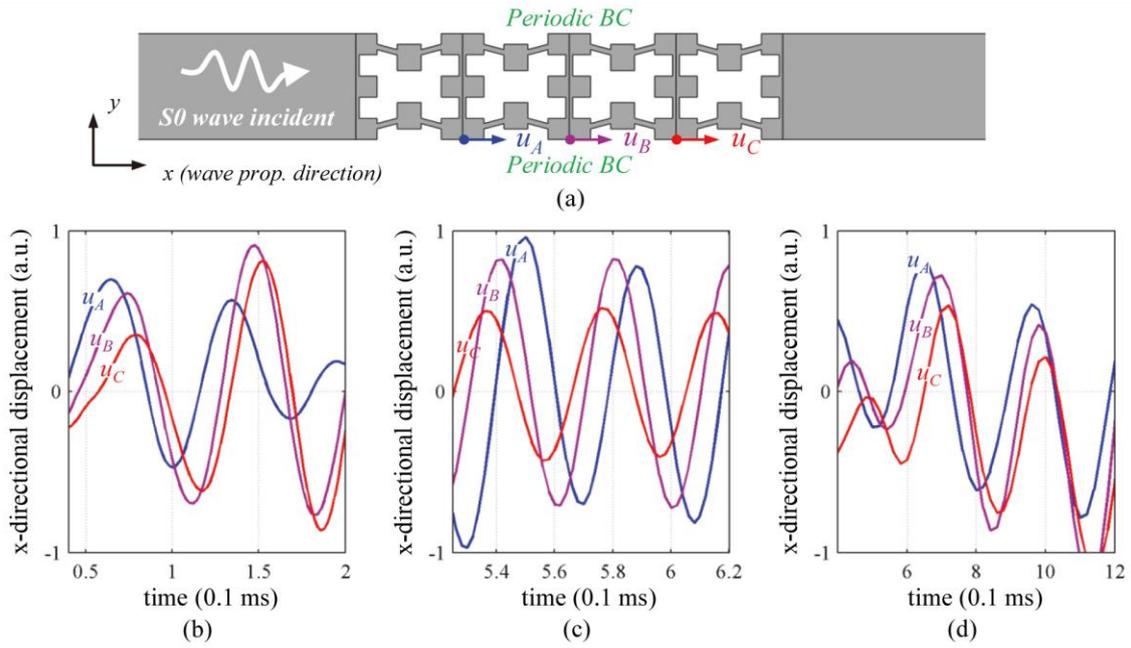

**Fig. S7.** (a) Numerical simulation modeling for the metamaterials made of $C_{mk}$, and the measured displacement values at the frequency of (b) 15, (c) 25 and (d) 35 kHz.

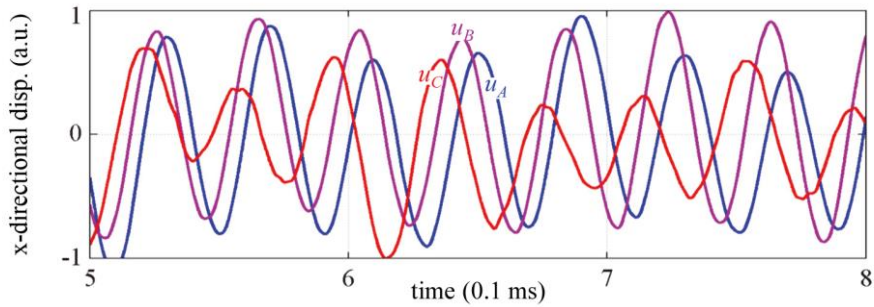

**Fig. S8.** Experimentally measured *x*-displacements ($u_A$, $u_B$ and $u_C$) by a laser vibrometer

**Experimental setting for the metamaterials**

Fig. S9 illustrates the schematic figure of the experimental setup. The imbedded metamaterials in the base aluminum plate are fabricated by the waterjet cutting process. The dimensions of the base plate are 2 m in width, 1.2 m in height and 1 mm in thickness. The imbedded metamaterial system fabricated in the middle of the base plate consists of $4 \times 25$ unit cells with 4 unit cells along the propagating $x$ direction. Fig. S10 shows the photos of the fabricated metamaterials made of $C_m$, $C_k$ and $C_{mk}$. The traction-free top and bottom surfaces ensure the formation of guided waves along the *x* direction[S7]. The actuator and sensor locations for wave experiments are shown in Fig. S9. Patch-type piezoelectric transducers (thickness: 1 mm, radius: 1.2 cm), as illustrated in Fig. S11 (a), were installed 5 cm away from the boundaries of the metamaterial system to generate and measure the S0 wave mode.

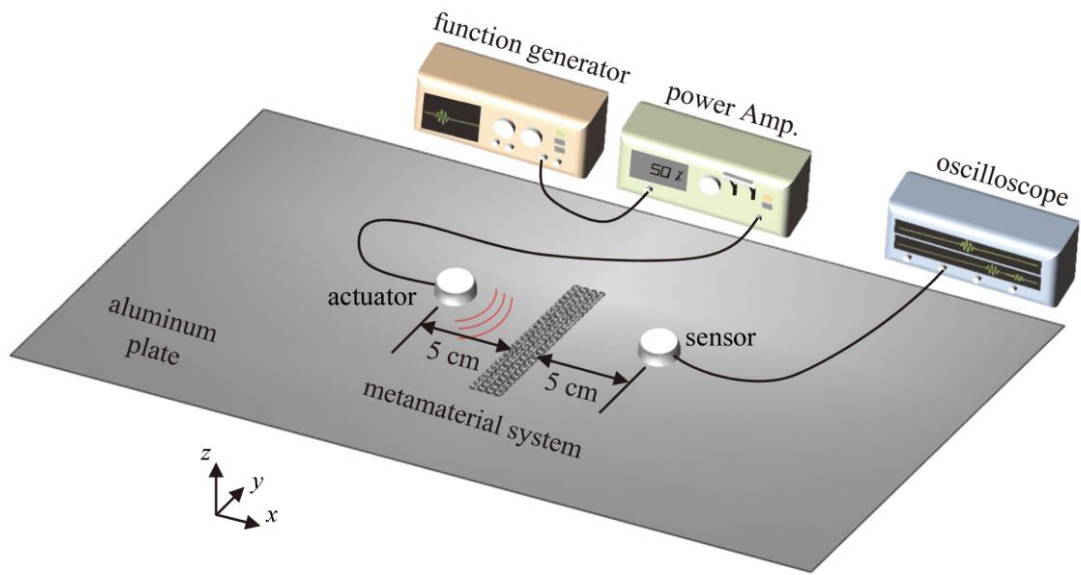

**Fig. S9.** Schematic figure of the experimental setup.

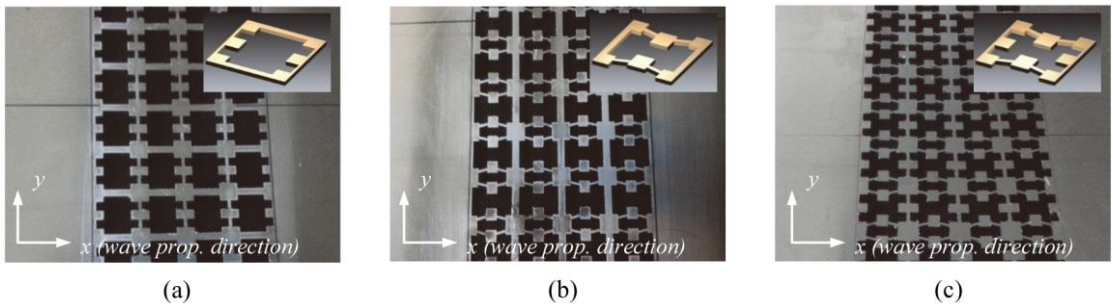

**Fig. S10.** Photos of the fabricated metamaterials made of (a) $C_m$ (b) $C_k$ and (c) $C_{mk}$, respectively. The experimental results in Fig. 5 are based on the fabricated metamaterials illustrated here.

To make sure that the dominant wave mode generated and measured with the used patch-type piezoelectric transducers is the S0 wave mode, not the undesirable A0 mode (the lowest anti-symmetric Lamb wave) in the frequency range of interest, a reference pitch-catch experiment was performed in a homogeneous aluminum plate. The input signal to actuate the patch-type piezoelectric transducer is shown in Fig. S11 (b) and the measured output signal by a receiving patch-type transducer that is 30 cm apart from the actuating transducer is plotted in Fig. S11 (c). Because the transducer is so configured as to predominantly generate the S0 wave mode[S8] and the group velocity of the S0 wave mode in the aluminum plate ($v_g^{al}\big|_{S0} \approx 5200$ m/s) is much faster than the group velocity of the A0 wave mode in the aluminum plate ($750 \leq v_g^{al}\big|_{A0} \leq 1150$ m/s in the frequency range of interest), we can use the first arrival signal of the S0 wave mode which can be completely distinguishable from the A0 wave mode. Inside the metamaterial, the difference in the group velocity between the S0 and A0 wave modes is more severe, which makes it easier to work only with the S0 wave mode. It was found that $v_g^{meta}\big|_{S0} \approx 2140$ m/s while $v_g^{meta}\big|_{A0} \approx 137$ m/s. Also, since the A0 wave mode of the metamaterial made of $C_{mk}$ has a stop band at around 25 kHz, only the S0 wave mode is measured at the double negative region of the metamaterial made of $C_{mk}$. Because we can capture the first arrival pulse of the S0 wave mode only in actual experiments, the resulting data analysis is performed with the S0 mode.

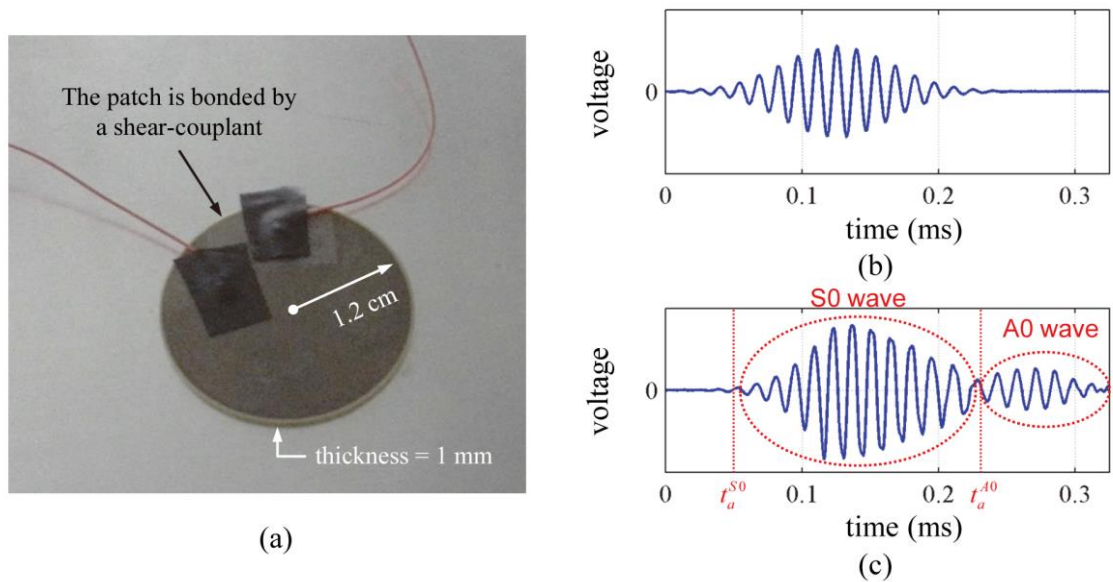

**Fig. S11.** (a) The piezoelectric transducer installed on an aluminum plate. (Note that a heavy mass is placed on it during experiments.) (b) Actuated input signal and (c) measured elastic wave signal from the transducer installed on a 1mm thick homogeneous aluminum plate.

The detailed experimental procedure is as follows. First, the actuation signal was generated by a function generator (33250A, Agilent Technologies Inc., Santa Clara, CA). The modulated Gaussian pulse, or the Gabor pulse, was used as the actuation signal which was generated by

$$\exp(-\frac{f^2}{2G_s^2}t^2)\cos(2\pi ft). \qquad (S27)$$

In equation (S27), $G_s$ is a factor that controls the time spread of the pulse and $f$ is center frequency of the pulse. In our experiments, $G_s$ was set to be 2.75 and the center frequencies ($f$) were chosen to be 15, 25 and 35 kHz. The actuated signal was amplified by a power amplifier (AG1017L, T&C Power Conversion, Rochester, NY) and sent to the actuating piezoelectric transducer. The wave transmitted through the metamaterial layer system was measured by the receiving piezoelectric transducer and the signal from the receiving transducer was recorded by an oscilloscope (WaveRunner 104MXi-A, LeCroy, Chestnut Ridge, NY). The excitation and measured signals at 35 kHz are shown in Figs. S12 (a,b). Measured signals were post-processed before they were compared with simulation results. For the comparison, we used the arrival time $t_a$ defined as

$$t_a = \frac{d^{al}}{v_g^{al}\big|_{S0}} + \frac{4d_{unit}^{meta}}{v_g^{meta}\big|_{S0}} = \frac{0.1}{v_g^{al}\big|_{S0}} + \frac{0.08}{v_g^{meta}\big|_{S0}} \tag{S28}$$

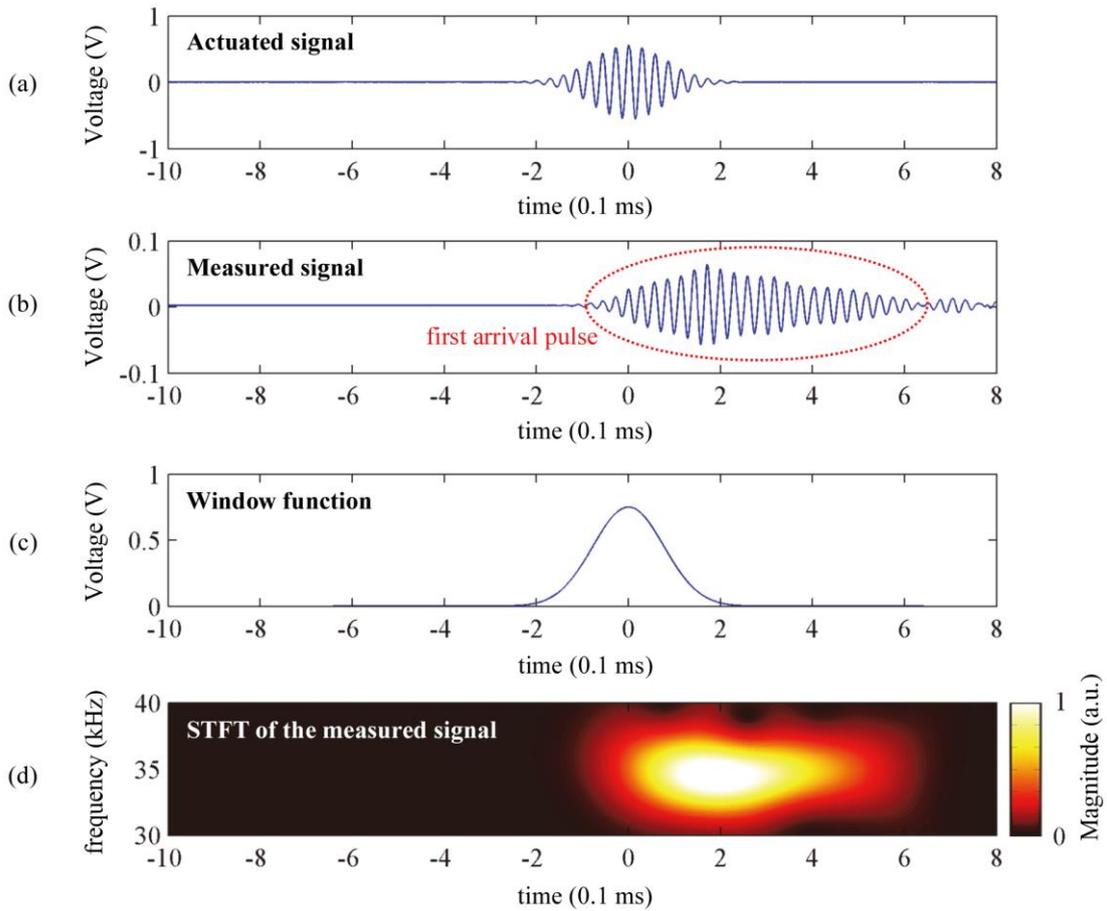

**Fig. S12.** Experimental signals obtained for the excitation frequency of $f = 35$ kHz. (a) Actuated signal, (b) measured signal, (c) the Gaussian window function used for the STFT and (c) the STFT of the first arrival pulse in (b).

where the superscripts '*al*' and '*meta*' stand for aluminum and metamaterial. The symbol $d_{al}$ denotes the sum of the distances from the actuating and receiving transducers to the metamaterial boundaries. The symbol $d_{unit}^{meta}$ represents the size of the metamaterial unit cell in the *x* direction. The experimental arrival time $t_a$ can be estimated from the experimental data by performing the short-time Fourier transformation (STFT) (see, e.g., Mallat[S9]) of the first arrival pulse (as marked in Fig. S12 (b)). The short-time Fourier transform $SV(\tau,\omega)$ of $V(t)$ is defined as

$$SV(\tau,\omega) = \int_{-\infty}^{\infty} V(t) g(t-\tau) \exp(-i\omega t) dt \quad \text{(S29)}$$

where $g(t)$ is a real symmetric window function, $\tau$ denotes the amount of the translation of $g(t)$ in time and $\omega$, the angular modulation frequency. The selected window function $g(t)$ is the Gaussian window plotted in Fig. S12 (c). Fig. S12 (d) plots the absolute value $|SV(\tau,\omega)|$ for varying $\tau$ (representing the horizontal axis) and $\omega$ (representing the vertical axis). Note that $|SV(\tau,\omega)|^2$ is usually called the spectrogram denoting the energy of $V(t)$ in the time-frequency neighbor of $(\tau,\omega)$. The level of color at $(\tau,\omega)$ in Fig. S12 (d) corresponds to the magnitude of the output voltage in the neighbor of $(\tau,\omega)$. If ridges are identified from $|SV(\tau,\omega)|$ or $|SV(\tau,\omega)|^2$, the information of the arrival time of a specific harmonic component can be extracted. (The analysis of dispersive waves by the STFT may be found in Ref. S10.) For instance, $t_a$ of the S0 wave can be found by reading the local maxima in the STFT plot in Fig. S12 (d). On the other hand, one can numerically calculate the arrival time $t_a$ by using the group velocity of the S0 wave mode $v_g^{meta}\big|_{S0}$ in the metamaterial and that in the base aluminum plate, $v_g^{al}\big|_{S0}$.

**Verification of independent tunability: Experiments and findings**

To verify the independent tunability of the proposed elastic metamaterial, additional experiments were performed. Starting from the original $C_{mk}$ configuration, the *x*- and *y*-resonating parts are separately varied to make $C_{\tilde{m}k}$ and $C_{m\tilde{k}}$ unit cells. They are compared against $C_{mk}$ in Table S2. The objective to perform experiments with $C_{\tilde{m}k}$ and $C_{m\tilde{k}}$ is to show that the negative density or the negative stiffness can be altered alone without affecting its counterpart, the negative stiffness or the negative density, respectively.

| Unit cell type | $C_{mk}$ (original) | $C_{\tilde{m}k}$ (change of the *x*-resonating part) | $C_{m\tilde{k}}$ (change of the *y*-resonating part) |
|---|---|---|---|
| Configuration | 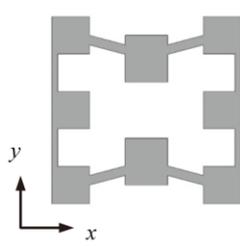 | 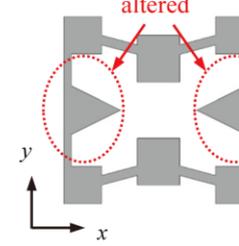 | 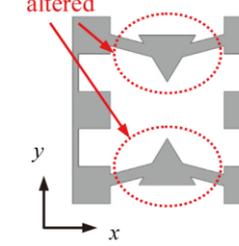 |
| $f_x$ Resonance freq. in *x* | 23.13 kHz | 32.19 kHz | 23.13 kHz |
| $f_y$ Resonance freq. in *y* | 30.08 kHz | 30.08 kHz | 38.59 kHz |
| Negative density range | 23.05 ~ 28.33 kHz | 32.30 ~ 37.25 kHz (almost identical with right) | 22.60 ~ 27.34 kHz |
| Negative stiffness range | 20.96 ~ 30.23 kHz (almost identical with middle) | 20.38 ~ 30.14 kHz | 30.87 ~ 38.66 kHz |

**Table S2.** Comparison of $C_{mk}$ and $C_{\tilde{m}k}$ against $C_{m\tilde{k}}$.

First, a new metamaterial made of $C_{\tilde{m}k}$ shown in Fig. S13 (a) is considered. In $C_{\tilde{m}k}$, the *x*-resonating part is replaced by a new resonator the resonance frequency of which is moved to 30.08 kHz while the *y*-resonating part is the same as that of $C_{mk}$. Therefore, two resonance gaps around 25 kHz (due to the negative stiffness) and 35 kHz (due to the negative density) are formed in the metamaterial made of $C_{\tilde{m}k}$. The experimental results in Fig. S13 (b) show that there is no S0 wave mode transmission around 25 and 35 kHz, indicating that the negative density region is moved to around 35 kHz. On the other hand, the negative stiffness region around 25 kHz is virtually unaltered; the range of the resonance gap due to the negative stiffness for the metamaterial made of $C_{\tilde{m}k}$ is 20.38 ~ 30.14 kHz while that for the metamaterial made of $C_{mk}$ is 20.96 ~ 30.23 kHz. This result confirms independent tuning of the negative density without affecting the negative stiffness of $C_{mk}$.

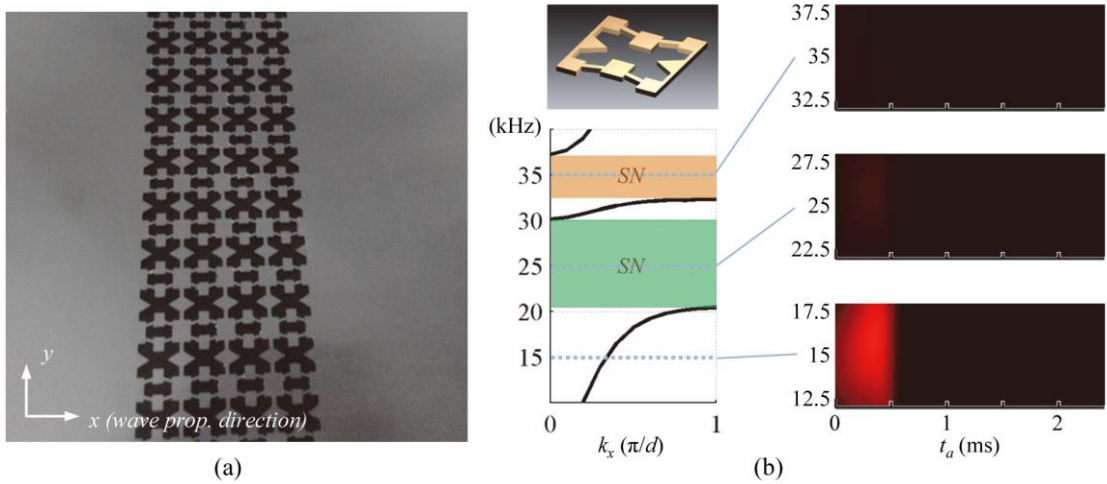

**Fig. S13**. (a) The picture of the metamaterial made of $C_{\tilde{m}k}$ for which only the *x*-resonance frequency is changed to around 32.19 kHz, (b) the dispersion curve and experimental results of the metamaterial made of $C_{\tilde{m}k}$.

Now, let us consider a new metamaterial made of $C_{m\tilde{k}}$ shown in Fig. S14(a). In this case, the resonance gap due to the negative stiffness is formed around 35 kHz while the resonance gap due to the negative density remains unaltered in comparison with the metamaterial made of $C_{mk}$; see Fig. S14 (b). The experimental results confirm that the range of the resonance gap due to the negative density for the metamaterial made of $C_{m\tilde{k}}$ is 22.60 ~ 27.34 kHz while that for the metamaterial made of $C_{mk}$ is 23.05 ~ 28.33 kHz. This result confirms independent tuning of the negative stiffness without altering the negative density of $C_{mk}$. From the experimental results in Figs. S13 and S14, negative density and stiffness are found to be independently tunable.

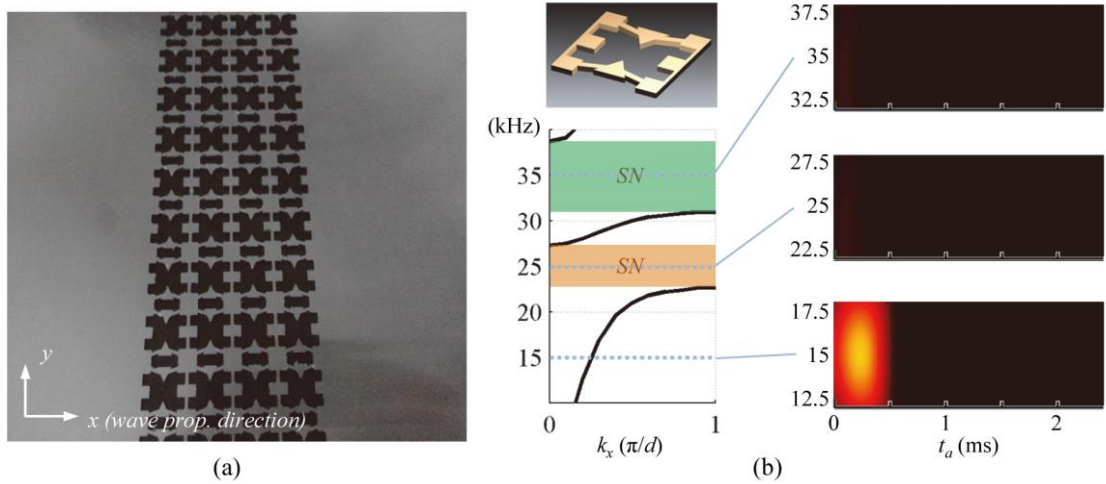

**Fig. S14**. (a) The picture of the metamaterial made of $C_{m\tilde{k}}$ for which only the *y*-resonance frequency is changed to 38.59 kHz, (b) the dispersion curve and experimental results of the metamaterial made of $C_{m\tilde{k}}$.